# Long-time traces of the initial condition in relaxation phenomena near criticality

U. Ritschel and H. W. Diehl

*Fachbereich Physik, Universität GH Essen, D-45117 Essen*

(September 16, 1994)

## Abstract

The time evolution of systems relaxing towards thermal equilibrium is examined near the critical temperature $T_c$, with special attention paid to the role of the initial value $m_i$ of the order parameter $\phi$. To this end, the $n$-component model A for a cube of length $L$ is investigated. The common belief that all memory of $m_i$ is necessarily lost after a microscopic time span is shown to be unfounded. General arguments and the exact solution of the limit $n \to \infty$ show that $m_i$ leaves its traces in both the linear and nonlinear long-time relaxation of $\phi$ near or at $T_c$. Specifically for linear relaxation near $T_c$, or at $T_c$ with $L < \infty$, the amplitude of the exponential decay depends on $m_i$ and the short-time exponent $\theta' = (x_i - x_\phi)/z$, provided $t_i \sim m_i^{-z/x_i}$ is comparable to or larger than other time scales. Here $x_i$ is the scaling dimension of $m_i$, $z$ is the dynamic bulk exponent, and $x_\phi$ is the usual equilibrium scaling dimension of $\phi$.

Typeset using REVTEX

How does a thermodynamic property of a system at temperature $T$ above, but near its critical point $T_c$, relax to zero with time $t$? This is a long-studied problem of critical dynamics, which we wish to re-examine here. The main question we wish to answer is whether and to what extent *the initial condition* affects such relaxation processes *on long time scales*.

The stochastic models used in the theory of dynamic critical phenomena [1] to represent the various dynamic universality classes, by construction, possess a number of properties such as detailed balance which guarantee that the system relaxes from almost any initial configuration towards the thermal equilibrium state. The common belief is that all memory of the initial state gets lost on a *microscopic* time scale. As we shall show, this is not in general the case: Under appropriate (realizable) conditions (to be discussed below), the initial condition may well leave its fingerprints on the relaxational behavior on long time scales.

To become more specific, let us consider the following relaxation process: A ferromagnetic system with non-conserved $n$-component order parameter $\phi$ and purely relaxational dynamics (model A of Ref. [1]) is prepared in a macrostate with homogeneous magnetization $m_i$ at temperature $T_0 \gg T_c$ and magnetic field $h > 0$. At $t = 0$ the magnet is rapidly quenched to a temperature $T \gtrsim T_c$ and $h$ is switched off. Then the system is let to evolve.

Immediately after the quench, up to some *microscopic* time scale $t_{\text{mic}}$, the time evolution clearly should be *non-universal*, depending on both the *microscopic* details of the interactions in the sample and the initial condition. For $t \to \infty$, the same unique equilibrium state is approached, irrespective of the precise initial condition. This suggests that for sufficiently large $t \gg t_{\text{mic}}$ the process should enter a regime of *universal* behavior, involving only a few macroscopic time scales.

One well-known such scale is the characteristic time $t_\tau \sim \tau^{-\nu z}$, where $\tau = (T - T_c)/T_c$; in our case it may be defined as the *linear relaxation time* [2] of the order parameter $m(t) = \langle \phi(\mathbf{x}, t) \rangle$ in a bulk system. For a long time it had been believed that only two time regimes with universal time dependence could be distinguished [1]: a regime $t \gg t_\tau$ of *linear relaxation*, $m(t) \sim e^{-t/t_\tau}$, and a regime $t_{\mathrm{mic}} \ll t \ll t_\tau$ of *nonlinear relaxation* [3,4] with $m(t) \sim t^{-\beta/\nu z}$. The problem with this simple picture is that it precludes the possibility of any initial-time effects on macroscopic time scales. This is unjustified since the initial value $m_i$ yields an independent time scale $t_i$ that may become arbitrarily large for sufficiently small $m_i$. In fact, if we assume that $m_i$ scales with a dimension $x_i$ under dilatations, then $t_i \sim m_i^{-z/x_i}$. This suggests that the above simple picture is incomplete and needs to be corrected.

A crucial step in this direction was made by Janssen et al. [5,6] who showed that, contrary to previous assumptions, $x_i$ is a *genuine new* exponent, generally different from the scaling dimension $x_\phi$ of $\phi(t>0)$. As a consequence, $m(t)$ was found to exhibit a non-trivial short-time behavior of the form $m(t) \sim t^{\theta'}$ for $t_{\mathrm{mic}} \ll t \ll t_i \ll t_\tau$, with $\theta' = (x_i - x_\phi)/z$. For model A below its upper critical dimension $d^* = 4$, $x_i$ turned out to be larger than $x_\phi$, so that $m(t)$ actually increases in this regime.

Ref. [5] and subsequent work [7] mainly dealt with the critical case ($t_\tau = \infty$). To complement these findings we here focus our attention on regimes in which $t \gg t_{\mathrm{mic}}$ is large compared to the linear relaxation time. While our main interest is in the bulk case, we will also discuss the finite-size case of a cube of length $L$ with periodic boundary conditions, so that the special case $t_\tau = \infty$, $L < \infty$, treated in Ref. [8] is included. For this special case we have shown there that the initial condition also affects the linear relaxational behavior on long time scales. Using general scaling arguments, we show below that this phenomenon is not restricted to finite-size systems at $T_c$ but can be observed *quite generally* in the case of *linear* relaxation. Our main findings will be corroborated by exact solutions for the $n \to \infty$ limit of the mentioned $n$-component model A with $L \leq \infty$.

To develop our argument we begin by recalling the scaling form [8]

$$m(t, \tau, L, m_i) \approx C_m \, t^{-\beta/\nu z} \, \mathcal{F}(t/t_\tau, t/t_L, t/t_i) \qquad (1)$$

phenomenological scaling considerations lead one to anticipate on large length and time scales. Here $C_m$ is a non-universal amplitude. Further non-universal metric factors appear in $t_\tau$, $t_L$, and $t_i$. The function $\mathcal{F}$ is universal.

In Ref. [8], the form of $\mathcal{F}(0, \vartheta_L, \vartheta)$ in various limits was discussed. For example, the anomalous short-time behavior $\sim t^{\theta'}$ could be traced back to the fact that $\mathcal{F}(0, 0, \vartheta)$ exists and behaves as

$$\mathcal{F}(0,0,\vartheta) \approx \mathcal{F}_{bci}\, \vartheta^{x_i/z} \tag{2}$$

as $\vartheta \equiv t/t_i \to 0$ [9], where $\mathcal{F}_{bci}$ is a universal constant.

In the long-time limit, with which we are concerned here, $m(t)$ must decay exponentially, unless $t_\tau = t_L = \infty$. That is, as $t \to \infty$ with fixed $t_\tau$, $t_L$, and $t_i$, it should vary as

$$m(t, t_\tau, t_L, t_i)/C_m \approx M_\infty\, e^{-t/t_E} \,. \tag{3}$$

Compatibility with (2) requires that the amplitude $M_\infty$ and the linear relaxation time $t_E$ have appropriate scaling forms. The former in general depends on $\tau$, $L$, and $m_i$; it can be written as

$$M_\infty(\tau, L, m_i) = t_\tau^{-\beta/\nu z}\, \mathcal{A}(t_L/t_\tau, t_i/t_\tau) \,. \tag{4}$$

On the other hand, $t_E$ should be independent [11] of $m_i$ and satisfy the relation

$$t_E(\tau, L) = t_\tau\, \mathcal{T}(t_L/t_\tau) \,. \tag{5}$$

Since $t_\tau = t_E(\tau, \infty)$ and $t_L = t_E(0, L)$, the function $\mathcal{T}(x)$ must have the following asymptotic properties

$$\mathcal{T}(x \to \infty) \approx 1 \quad \text{and} \quad \mathcal{T}(x \to 0) \approx x \,. \tag{6}$$

Turning to the universal amplitude function $\mathcal{A}(x, y)$, we first note that the bulk limit $x \to \infty$ of (4), and hence

$$\mathcal{A}(x \to \infty, y) \approx \mathcal{A}_b(y) \,, \tag{7}$$

should exist. At criticality, (4) should become

$$M_\infty \approx t_L^{-\beta/\nu z} \mathcal{B}_c(t_i/t_L) \ . \tag{8}$$

This implies the behavior

$$\mathcal{A}(x, w/x) \approx x^{-\beta/\nu z} \mathcal{B}_c(w) \tag{9}$$

as $x \to 0$ with fixed $w = t_i/t_L$. If both $t_\tau$ and $t_L$ are large compared to $t_i$, it seems reasonable to expect (in complete accordance with our exact results below) that any dependence on $m_i$ should drop out of the limiting form of $m(t)$ and hence of $M_\infty$. From this we infer the asymptotic properties

$$\mathcal{A}(x, 0) = \mathcal{A}_\infty(x) \quad \text{and} \quad \mathcal{B}_c(0) = \mathcal{B}_{c\infty} \ , \tag{10}$$

in which $\mathcal{B}_{c\infty} \neq 0$ is a finite universal number.

Of particular interest are the cases with $t_i$ large compared to $t_\tau$, to $t_L$, or to both. For those our exact results below and in Ref. [8] reveal a dependence of $M_\infty$ on $m_i$, suggesting that this dependence should be linear. In order that this behavior complies with (4) we must have

$$\mathcal{A}(x, y \to \infty) \approx y^{-x_i/z} \mathcal{A}_i(x) \ . \tag{11}$$

From this we immediately deduce the limiting form

$$M_\infty \approx \text{const } m_i \, t_\tau^{\theta'} \mathcal{A}_i(t_L/t_\tau) \tag{12}$$

for $t_i \to \infty$ with fixed $t_\tau$ and $t_L$. This is a central result of this Letter. It shows that the amplitude $M_\infty$ depends indeed on the initial value $m_i$, and that *the short-time exponent $\theta'$ governs the $\tau$-dependence of $M_\infty$*. For the function $\mathcal{A}_i$ we anticipate the limiting properties

$$\mathcal{A}_i(x \to \infty) \approx \mathcal{A}_{ib} \quad \text{and} \quad \mathcal{A}_i(x \to 0) \approx \mathcal{A}_{ic} \, x^{\theta'} \ . \tag{13}$$

The first ensures that both above features survive the bulk limit $L \to \infty$; the second implies that (12) reduces for $t_\tau \gg t_i \gg t_L$ correctly to the $\tau = 0$ form

$$M_\infty \approx \text{const}\, m_i\, t_L^{\theta'}\, \mathcal{A}_{ic} \tag{14}$$

discovered in Ref. [8].

The scaling forms predicted above may all be verified explicitly in the limit $n \to \infty$ of our model. This is defined by the Langevin equation

$$\lambda^{-1} \partial_t \phi(\mathbf{x}, t) = \left(-\Delta + \tau + \tau_c + \frac{g}{6n} \phi^2\right) \phi + \zeta, \tag{15}$$

in which $\tau_c$ is the usual static critical value of the bare squared mass, while $\zeta$ is a Gaussian random force with zero mean and variance

$$\langle \zeta_\alpha(\mathbf{x}, t)\, \zeta_\beta(\mathbf{x}', t') \rangle = 2\, \delta_{\alpha\beta}\, \delta(\mathbf{x} - \mathbf{x}')\, \delta(t - t') \tag{16}$$

We assume that $2 < d < 4$, so that an ordered equilibrium state exists and hyperscaling is valid. In the limit $n \to \infty$ the model becomes Gaussian with a time-dependent susceptibility $\chi(t)$; its dynamics is completely described by the self-consistent set of equations

$$\chi(t)\, \partial_t m(t) = -\lambda\, m(t) \tag{17}$$

and

$$\chi(t)^{-1} = \tau + \tau_c + \frac{g}{6} \left[C(t) + m(t)^2\right] \tag{18}$$

with

$$C(t) = \sum_{\mathbf{q}} C(\mathbf{q}; t, t), \tag{19}$$

where $m(t)$ now means the rescaled magnetization $n^{-1/2} |\langle \phi(\mathbf{x}, t) \rangle|$ with initial value $m(t = 0) = m_i$. For brevity all arguments of $m(t, \tau, L, m_i)$ but $t$ are omitted. The sum in (19) extends over discrete momenta $\mathbf{q} = 2\pi\, \mathbf{m}/L$ with $\mathbf{m} \in \mathbb{Z}^d$. The Fourier-transformed autocorrelation function $C(\mathbf{q}; t, t)$ may be expressed in terms of the response propagator

$$G(\mathbf{q}; t, t') = \theta(t - t')\, \exp\left(-\lambda \int_{t'}^{t} dt'' \left[\mathbf{q}^2 + \chi(t'')^{-1}\right]\right) \tag{20}$$

as

$$C(\mathbf{q};t,t) = 2\lambda \int_0^\infty dt' \, G(\mathbf{q};t,t') \, G(\mathbf{q};t,t') \,. \tag{21}$$

When (17) and (18) are combined, $\chi(t)$ can be eliminated to obtain a linear integro-differential equation for $f(t) \equiv 1/m^2(t)$ that can be solved by Laplace transformation [12]. The result for the Laplace transform $\tilde{f}(s) \equiv \int_0^\infty dt \, e^{-st} f(t)$ can be written down in explicit form. Ignoring corrections to scaling, we find

$$\tilde{f}(s) = (2\lambda \, A_d)^{-1} L^d \frac{D_g/L^2 \, m_i^2 + 2\lambda/sL^2}{h \left(sL^2/2\lambda\right) - D_g \, L^{d-2} \tau/A_d} \tag{22}$$

where

$$h(x) = x^{d/2-1} + \frac{1}{\Gamma(1-d/2)} \int_0^\infty d\xi \, \xi^{-d/2} k(\xi) \, e^{-x\xi} \tag{23}$$

with

$$k(\xi) = \left( \sum_{n=-\infty}^\infty e^{-n^2/4\xi} \right)^d - 1 \,, \tag{24}$$

$$A_d = -(4\pi)^{-d/2} \Gamma(1-d/2) \,, \quad \text{and} \quad D_g = 6/g \,.$$

The result can be cast in the expected scaling form

$$\tilde{f}(s) = \text{const} \, t_L^{1+2\beta/\nu z} \, \mathcal{G}\left(st_L, t_\tau/t_L, t_i/t_L\right) \,, \tag{25}$$

with $1 + 2\beta/\nu z = d/2$ for the $n = \infty$ case considered here. The relaxation times $t_\tau$ and $t_L$ [normalized in accordance with (3) and (5)] are explicitly given by

$$t_\tau = \frac{1}{\lambda} \left( \frac{D_g \tau}{A_d} \right)^{-\nu z} \quad \text{and} \quad t_L = \frac{L^z}{\lambda \, x_0} \tag{26}$$

with the $n = \infty$ values $\nu z = 2/(d-2)$ and $z = 2$. The parameter $x_0$ denotes the zero of $h(x)$. Upon taking $t_i$ as

$$t_i = \frac{D_g \, (d-2)}{4\lambda} \, m_i^{-z/x_i} \tag{27}$$

with the $n = \infty$ value $z/x_i = 2$, one can derive from (25) the scaling function

$$\mathcal{G}(u,v,w) = \frac{2w/(d-2) + 1/u}{h(x_0 u/2) - (x_0/v)^{(d-2)/2}} . \qquad (28)$$

For general values of $t_\tau$, $t_L$, and $t_i$ the Laplace back transform of (22) — and hence the scaling function $\mathcal{F}$ — cannot be calculated analytically but must be computed numerically [8]. However, $\mathcal{F}_{bc}(\vartheta) \equiv \mathcal{F}(0,0,\vartheta)$, the scaling function pertaining the bulk critical case, can be obtained in closed form. It reads

$$\mathcal{F}_{bc}(\vartheta) = \left(\frac{\vartheta}{\vartheta + 1}\right)^{1/2} . \qquad (29)$$

For $\vartheta \ll 1$, $\mathcal{F}_{bc} \approx \vartheta^{1/2}$, in accordance with (2). In the large-$\vartheta$ limit, $\mathcal{F}_{bc} = 1 - 1/2\vartheta + O(\vartheta^{-2})$, i.e., the leading term is *independent* of $\vartheta$. This bears out that the resulting asymptotic long-time form of $m(t)$, the well-known nonlinear relaxation, becomes indeed independent of the initial value $m_i$.

The asymptotic long-time behavior (3) of $m(t)$ in the case $t_\tau + t_L > 0$, and the scaling function $\mathcal{A}$ of (4), can also be obtained analytically, employing standard results about Laplace transformation [13]. The asymptotic exponential decay follows from the contribution of the pole of $\tilde{f}(s)$ with largest positive real part, which in turn may be identified with $2/t_E$. One thus finds that the scaling function $\mathcal{T}(x = t_L/t_\tau)$ of (5) is the solution to

$$h(x_0 x/\mathcal{T}) - (x_0 x)^{(d-2)/2} = 0 . \qquad (30)$$

Noting that $h(x_0) = 0$ and $h(x \to \infty) \approx x^{d/2-1}$, one easily verifies that the limiting properties (6) are satisfied.

The result for the scaling function $\mathcal{A}(x,y)$ reads

$$\mathcal{A}(x,y) = (x_0\, x)^{\theta'} \left(\frac{2\, h'(x_0 x/\mathcal{T})}{(d-2)\mathcal{T} + 4\, y}\right)^{1/2} \qquad (31)$$

with the $n = \infty$ exponent value $\theta' = (4-d)/4$ and where we have chosen the normalization such that $\mathcal{A}(x \to \infty, y \to 0) \to 1$. Our result bears out the fact that the amplitude $M_\infty$ defined in (3), in general, depends on $t_i$ and hence on $m_i$. Memory of the initial value is lost only by decay.

Starting from (31) all the asymptotic cases discussed above can be studied in detail. Especially the scaling functions $\mathcal{A}_b$, $\mathcal{B}_c$, $\mathcal{A}_\infty$, and $\mathcal{A}_i$ introduced in (7), (9), (10), and (11), respectively, can be calculated explicitly. For $y = 0$ in (31), we obtain

$$\mathcal{A}_\infty(x) = (x_0\, x)^{\theta'} \left[\frac{2h'(x_0 x/\mathcal{T})}{(d-2)\,\mathcal{T}}\right]^{1/2} . \tag{32}$$

In this case the dependence on the initial field drops out, only subleading terms in the asymptotic expansion depend on the initial conditions.

In the complementary limit, $y \to \infty$, we identify by comparing with (11)

$$\mathcal{A}_i(x) \approx 2^{-1/2}\,(x_0\, x)^{\theta'}\,[h'(x_0 x/\mathcal{T})]^{1/2} , \tag{33}$$

and the limiting forms (12) and (14) can be verified. Hence, in the regime $t_i \gg t_\tau$, $t_L$ the amplitude $M_\infty$ indeed depends linearly on $m_i$. Further, the $\tau$ or $L$-dependence of $M_\infty$ for $t_i \gg t_L \gg t_\tau$ or $t_i \gg t_\tau \gg t_L$, respectively, is governed by the universal short-time exponent $\theta'$. For the latter case it is essential that $\theta'$ also occurs in the scaling function $\mathcal{A}(x,y)$.

So far we have focused our attention on the magnetization, the one-point function of the field theory. A natural question to ask is whether the above long-time dependence on $m_i$ also occurs in other quantities, in particular, in two-point functions. For our model with $n = \infty$ the result for the response propagator reads

$$G(\mathbf{q}; t, t') = \theta(t - t')\,\frac{m(t)}{m(t')}\, e^{-\lambda q^2\,(t-t')} . \tag{34}$$

Thus, in the $t, t' \to \infty$ limit it is given by

$$G(\mathbf{q}; t, t') \approx \theta(t - t')\, e^{-(\lambda q^2 + 1/t_E)(t-t')} , \tag{35}$$

that is all dependence on $m_i$ drops out. A similar result holds for the two-point cumulant.

A second obvious question is whether similar long-time traces will also occur for other models. Since our scaling arguments have been fairly general, we expect this to happen for those universality classes that have a non-trivial critical short-time behavior — i.e., for models with a non-vanishing short-time exponent $\theta'$. As shown in Ref. [7], interesting critical short-time behavior also occurs when (i) the order parameter is coupled to a conserved density (model C) and (ii) when one has reversible mode coupling (models G and E). It would certainly be interesting to extend our above analysis to these cases.

In summary, we have shown that, against the common belief, initial conditions affect critical relaxation processes even in the long-time regime. As expressed in (4) and explicitly verified for the large-$n$ limit, the amplitude of the linear decay is governed by the initial value of the order-parameter field as long as $t_i$, the initial time scale, is comparable or larger than other time scales. As critical dynamics is now in reach of computer simulations [14], it should be also possible to verify our results numerically.

*Acknowledgements:* We are grateful to the Deutsche Forschungsgemeinschaft for partial support through Sonderforschungsbereich 237.